\documentclass[preprint,12pt,authoryear]{elsarticle}

\usepackage{epsfig}

\usepackage{amssymb}
 \usepackage{amsthm}

 \usepackage{lineno}

\journal{Nuclear Physics A}

\begin{document}

\begin{frontmatter}



\title{Operation and performance of the ALICE Muon IDentifier RPCs during LHC Run3}


\author{Livia Terlizzi on behalf of the ALICE Collaboration}

\affiliation{organization={University and INFN Torino},%Department and Organization
            addressline={Via Pietro Giuria, 1}, 
            city={Torino},
            postcode={10125}, 
            state={Italia}}

\begin{abstract}
ALICE, which stands for ``A Large Ion Collider Experiment", is designed to study hadronic collisions at ultrarelativistic energies at the LHC. The primary objective of ALICE is to investigate the properties of quark-gluon plasma (QGP), a state of matter where quarks and gluons are deconfined under extreme conditions of temperature and energy density. One of the key observables for studying QGP is the production of hadrons carrying heavy quarks in Pb-Pb collisions. To detect heavy quarks via their muonic decays, ALICE is equipped with a forward muon spectrometer (MS). 
During LHC Run 1 and Run 2, a dedicated muon trigger system based on Resistive Plate Chambers (RPCs) was used for muon physics events selection. In preparation for Run 3, ALICE underwent a major upgrade, transitioning to a continuous readout (triggerless) mode to cope with higher interaction rates. As a result, the muon trigger system was replaced by the Muon Identifier (MID). In order to prevent ageing effects and to improve the RPC rate capability, the RPCs are now operated with lower gain by reducing the operating voltage while maintaining the same gas mixture. The front-end and readout electronics were also upgraded to support low-gain operation and triggerless readout. 
This paper presents an assessment of the stability and performance of the MID RPCs during the first two and a half years of Run 3 at unprecedented collision energies.
%During the LHC Run 1 and Run 2 the selection of interesting events for muon physics in ALICE was performed with a dedicated muon trigger system based on Resistive Plate Chambers operated in maxi-avalanche mode. During the long shutdown 2 (2019-2021) of LHC, ALICE conducted a major upgrade of its apparatus. The upgrade enables a new ambitious program of high-precision measurements. Moreover, in order to fully profit from the increased interaction rate to 50 kHz in Pb–Pb collisions (it was 10 kHz in Run 2) the ALICE experiment is running in continuous readout (triggerless) mode, hence the muon trigger became the Muon IDentifier (MID).
%In order to prevent ageing effects and to improve the RPC rate capability, it was decided to operate the detector with a lower gain, keeping the same gas mixture and decreasing significantly the working voltage. The front-end and readout electronics of the Muon Identification System have been upgraded in order to support low-gain operation and triggerless readout.
%The stability and performance of the MID RPCs during the first two and a half years of Run 3, at the unprecedented center-of-mass energies of 13.6 TeV for pp collisions and 5.36 TeV/nucleon pair for Pb–Pb collisions, are addressed in this paper.

\end{abstract}




\end{frontmatter}


%\section{The ALICE Muon Spectrometer (MS)}
%\label{ALICE_MS}

%The ALICE MS detects muons in the polar angular range 2° - 9°, i.e. it covers the pseudo-rapidity range 2.5 $< \eta <$ 4. It consists of: a silicon pixel sensors detector, based on Monolithic Active Pixel Sensors, called the Muon Forward Tracker (MFT) [4]; a front absorber to stop most hadrons emitted in the MS acceptance in order to minimize the residual hadron contamination (namely pions); a tracking system made of 5 stations of 2 planes of Cathode Pad and Cathode Strip Chambers, named Muon Chambers (MCH); a dipole magnet that provides a horizontal total field integral of 3 Tm and a maximum field of 0.7 T, perpendicular to the beam axis; a 1.2 m thick iron wall to filter the residual background of hadrons; finally there is the Muon IDentifier (MID) [5], described in the next section. \\

%\begin{figure}
%    \centering
%    \includegraphics[scale=0.4]{MS_MFT.png}
%    \caption{Schematic view of the ALICE Muon Spectrometer during Run 3}
%    \label{fig:MSRun3}
%\end{figure}

\section{The Muon IDentifier (MID)}
\label{MID_subsec}

The ALICE [1] MID [2] [3] consists of 72 Resistive Plate Chambers (RPCs) arranged in two stations, each containing two planes. Each plane measures $\approx$5.5 $\times$ 6.5 m$^2$, with a central hole of $\approx$ 1.2 $\times$ 1.2 m$^2$ to accommodate the beam pipe and its shielding. The area of a single RPC is $\approx $ 70 $\times$ 270 cm$^2$, and there are three different shapes: Long (L), Cut (C), and Short (S). The RPCs are equipped with orthogonal strips to provide spatial information along the X and Y directions, for a total of 21k strips with pitches of 1, 2, and 4 cm. The strip pitch increases with distance from the beam line to ensure a relatively uniform occupancy across the detector surface.

%\begin{figure}[htb]
%    \centering
%    \includegraphics[width=8.5cm,height=4cm]{ms_stazioni_layout.png}
%    \caption{Left: schematic view of the MID. Right: composition of an half plane. The different colors correspond to a different strip segmentation.}
%    \label{fig:MID_schem}
%\end{figure}

The MID RPCs are 2 mm single gap detectors with electrodes made of 2 mm thick high-pressure laminate (commonly referred to as bakelite), which has a resistivity $\rho$ in the range 3$\times$10$^9$ $\div$ 1$\times$10$^{10}$ $\Omega$cm.  The signal is inductively picked up using copper strips with an impedance of 50 $\Omega$. The gas mixture consists of 89.7$\%$ C$_2$H$_2$F$_4$, 0.3$\%$ SF$_6$, 10$\%$ \textit{i}-C$_4$H$_{10}$, humidified to 35-40$\%$ relative humidity. \\
Starting from Run 3, the detector operates in continuous readout mode (i.e. without a trigger), necessitating an upgrade of the readout electronics. To enhance the detector's rate capability and to prevent aging effects [4], the decision was made to operate the RPCs in avalanche mode rather than the so-callec maxi-avalanche mode used in Run 2, which had an average charge per hit of 100 pC. In this new operational mode, the RPCs function at a lower gain, with five times less charge released into the gas volume compared to the previous front-end electronics. This improvement was made possible by the newly introduced front-end electronics, based on the FEERIC ASIC [5], which includes a pre-amplification stage. Additionally, some RPCs accumulated a significant charge during Runs 1 and 2, potentially leading to sub-optimal performance due to aging effects. As a result, a new production of RPCs was launched, and four new detectors were installed in the cavern to replace those exhibiting large dark currents or gas leaks.

\section{MID Run 3 performance}
\label{MID_CERN}

The MID was fully operational when the first Run 3 stable beams at top energy began circulating in the LHC in July 2022. Since then, it has been running stably, with all the 72 RPCs functioning as expected.

\subsection{HV scan}

\begin{figure}[htb]
    \centering
    \includegraphics[width=14cm,height=9cm]{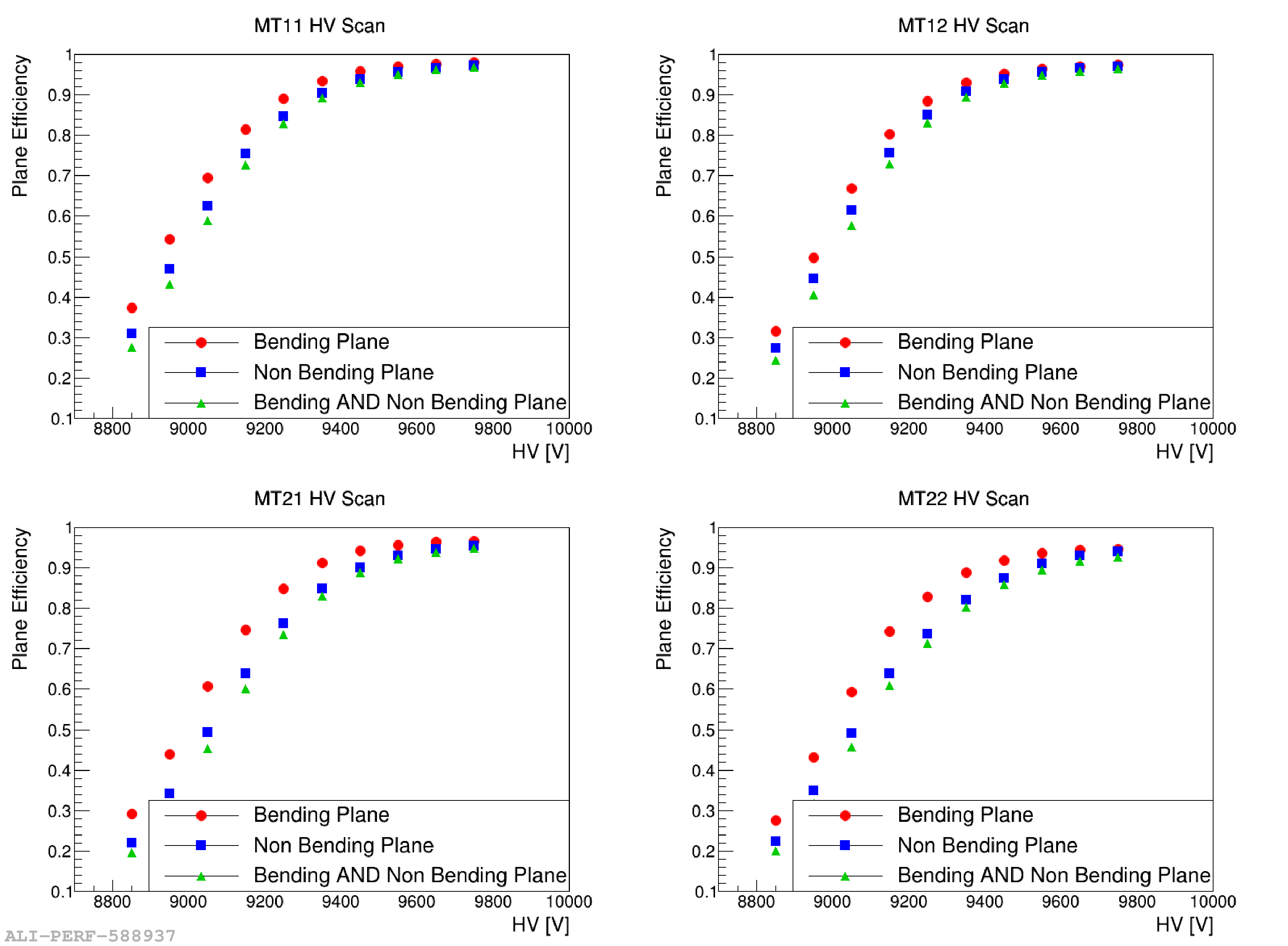}
    \caption{Efficiency as a function of HV for each of the four MID planes. In each plot three cases are shown: BP, NBP and both BP and NBP}
    \label{fig:MID_eff_planes}
\end{figure}

A fine tuning of the new working point (w.p.) high voltage (HV) values for each RPC was performed with p-p data at 500 kHz interaction rate. Before the HV scan, the RPCs were initially operated at a provisional w.p. of -700 V with respect to Run 2, corresponding to approximately 10300 V [6]. The scan was performed by varying the HV on one plane at a time, while the other three planes were kept at the provisional w.p. For each plane, 10 data sets were taken, with HV values ranging from -1400 V to – 500 V relative to the Run 2 working point, in steps of 100 V. The efficiency was evaluated at different levels: for the four planes, the 72 RPCs, and the 234 Local Boards (LBs), in three configurations, i.e. Bending Plane (BP) only, Non-Bending Plane (NBP) only, and both BP and NBP. The Bending Plane corresponds to the xy plane, while the Non-Bending Plane corresponds to the xz plane, with z representing the coordinate along the beam line. \\
The efficiency as a function of the average HV of the RPCs in the four MID planes is shown in Fig. \ref{fig:MID_eff_planes}. In each plot the BP, NBP and both BP and NBP cases are shown. The efficiency plateau is reached at 9700 V for all four planes, corresponding to -600 V relative to the Run 2 average HV working point.

\begin{figure}[hbt!]
\centering
\begin{minipage}{.45\linewidth}
  \centering
  \includegraphics[width=\linewidth]{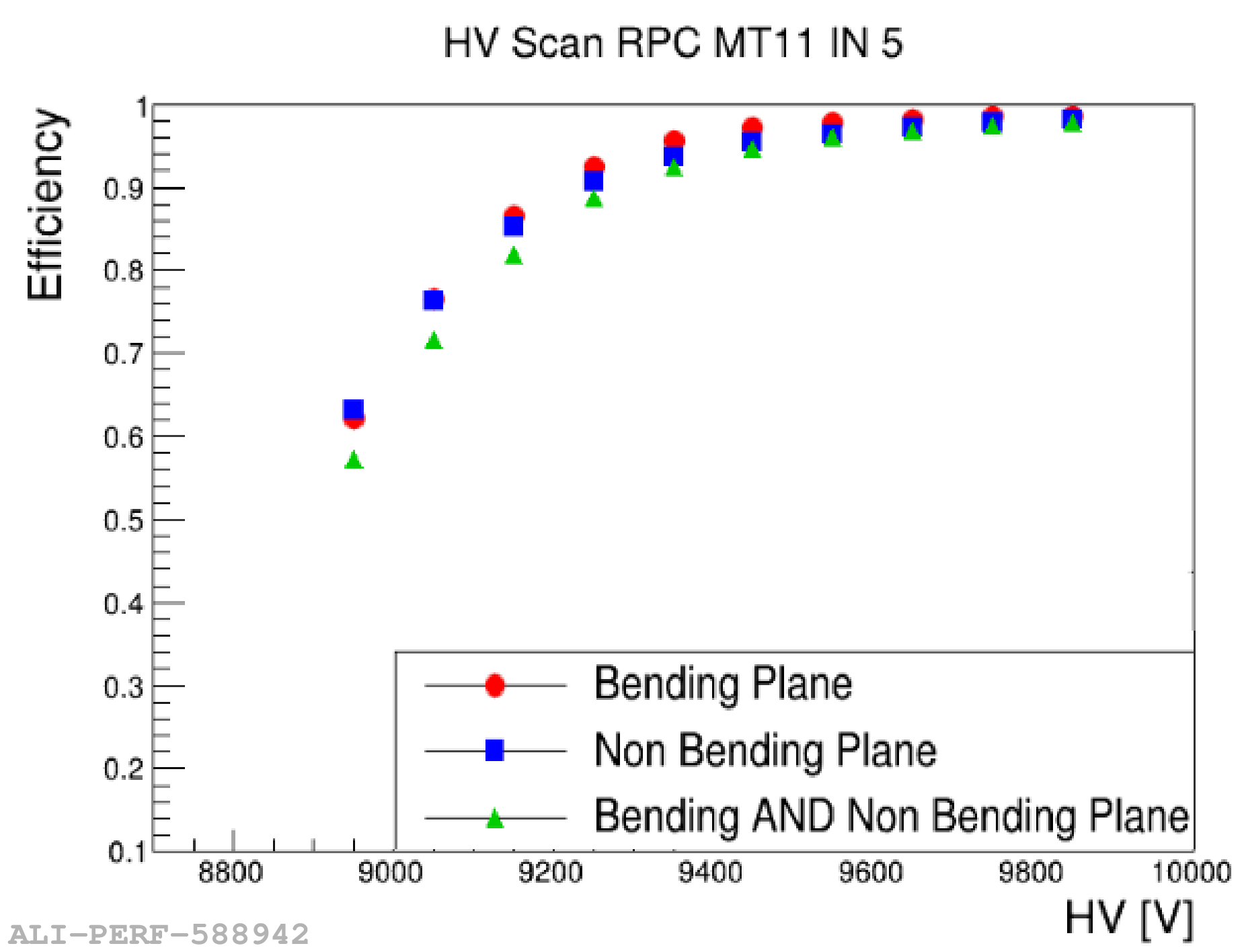}
\end{minipage}
\quad
\begin{minipage}{.45\linewidth}
  \centering
  \includegraphics[width=\linewidth]{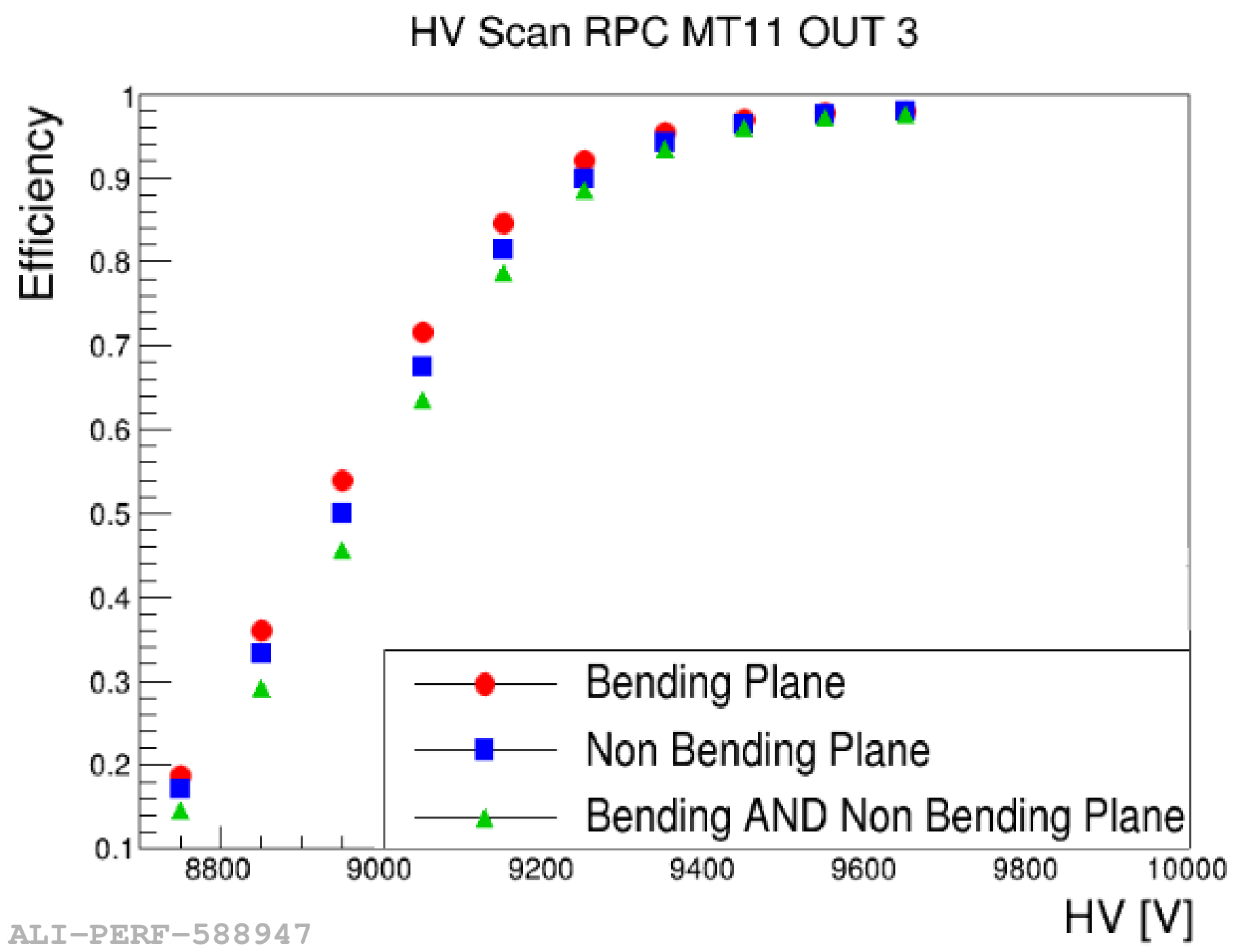}
\end{minipage}
\begin{minipage}{.45\linewidth}
  \centering
  \includegraphics[width=\linewidth]{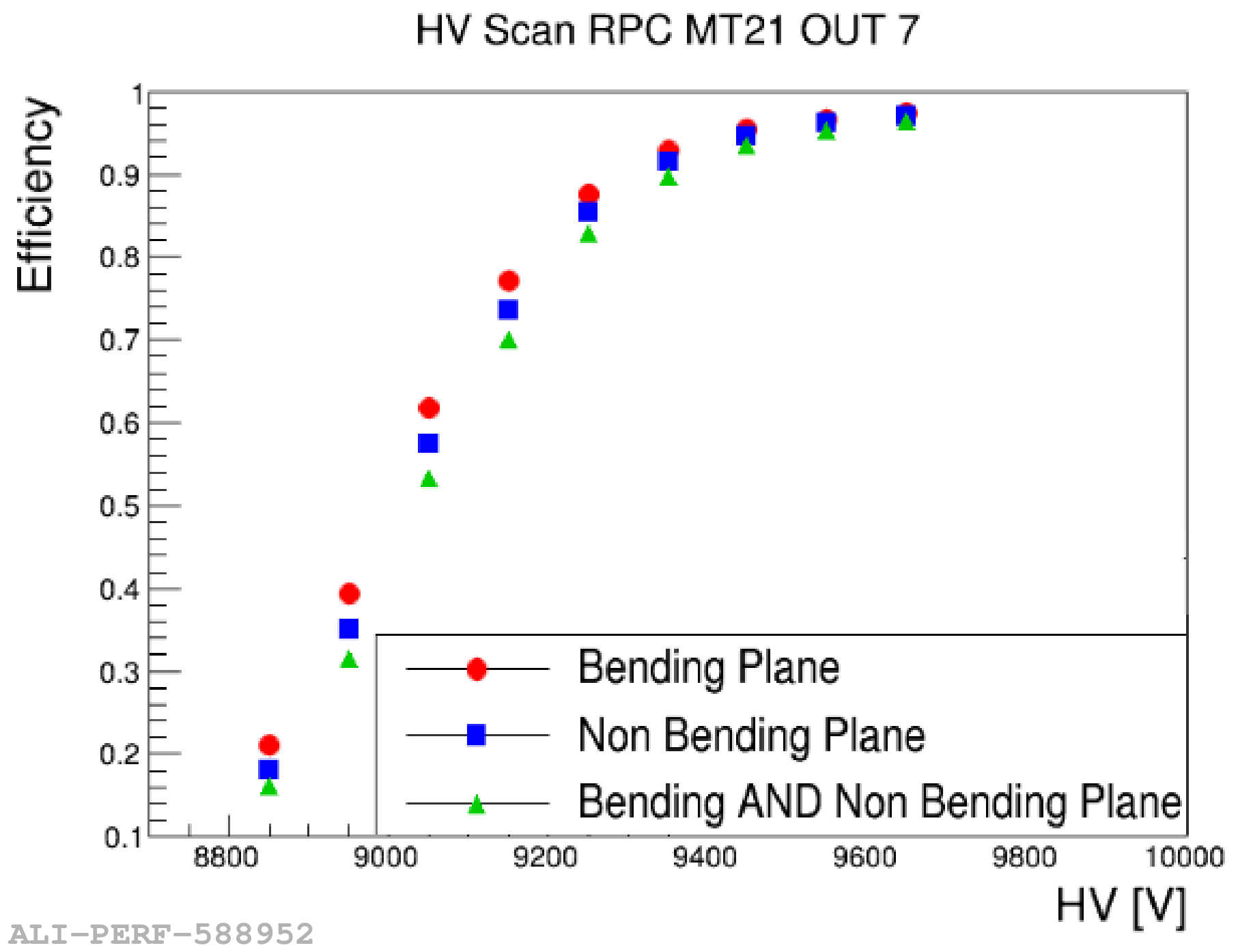}
\end{minipage}
\quad
\begin{minipage}{.45\linewidth}
  \centering
  \includegraphics[width=\linewidth]{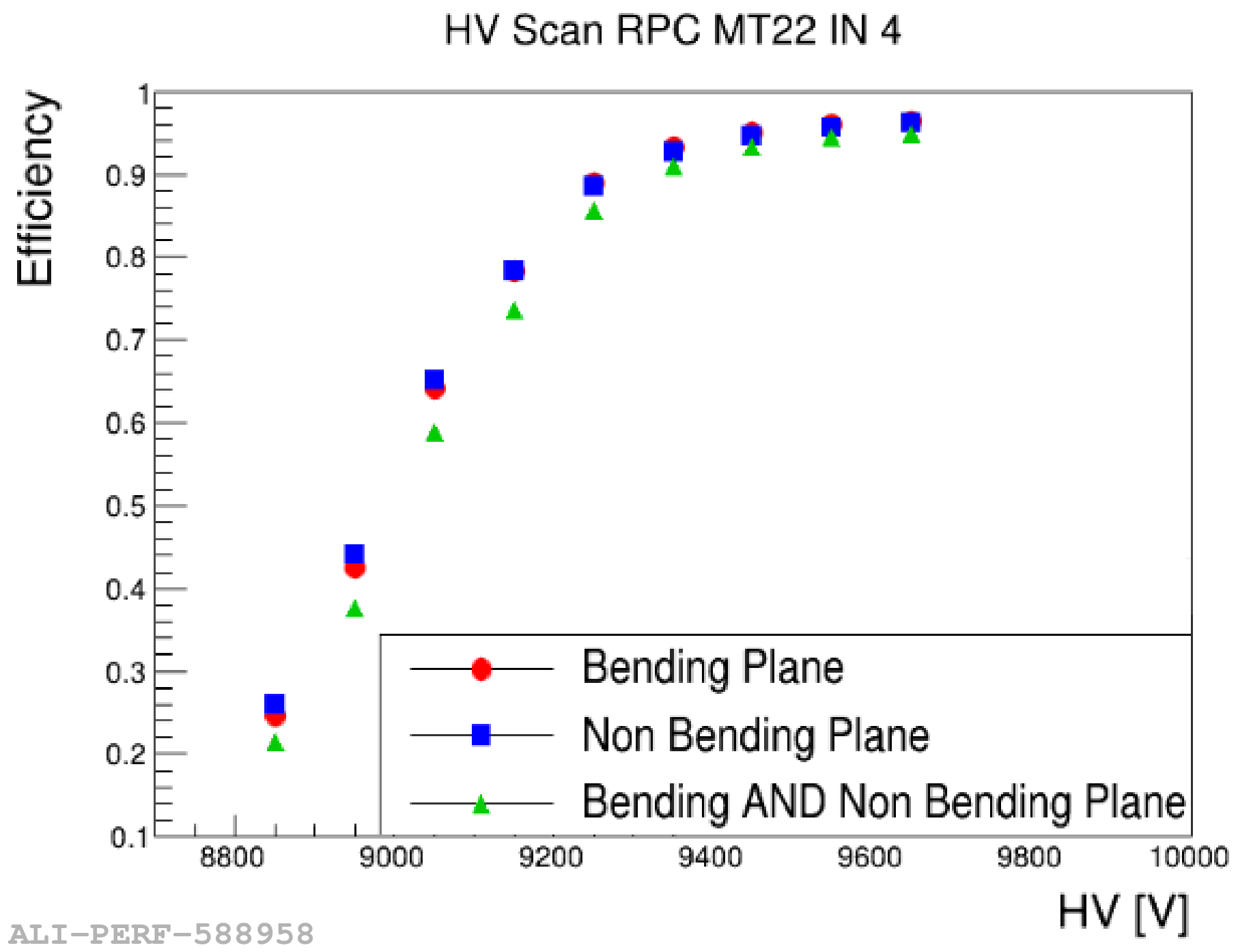}
\end{minipage}
\caption{Efficiency as a function of HV for 4 of the 72 MID RPCs. In each plot three cases are shown: BP, NBP and both BP and NBP}
\label{fig:MID_eff_RPC}
\end{figure}

In addition to the plane-level analysis, the efficiency was also evaluated at the individual RPC level to fine-tune the HV working point for each detector. The efficiency for 4 of the 72 MID RPCs is shown in Fig. \ref{fig:MID_eff_RPC}, as examples. While the majority of RPCs exhibit efficiency curves reaching the plateau, a few require further investigation due to slightly lower efficiency. Consequently, studies at the Local Board level are still ongoing. Similar to the plane-level results, the majority of detectors reach the plateau at approximately 9700 V.

\begin{figure}[htb]
    \centering
    \includegraphics[width=9cm,height=6cm]{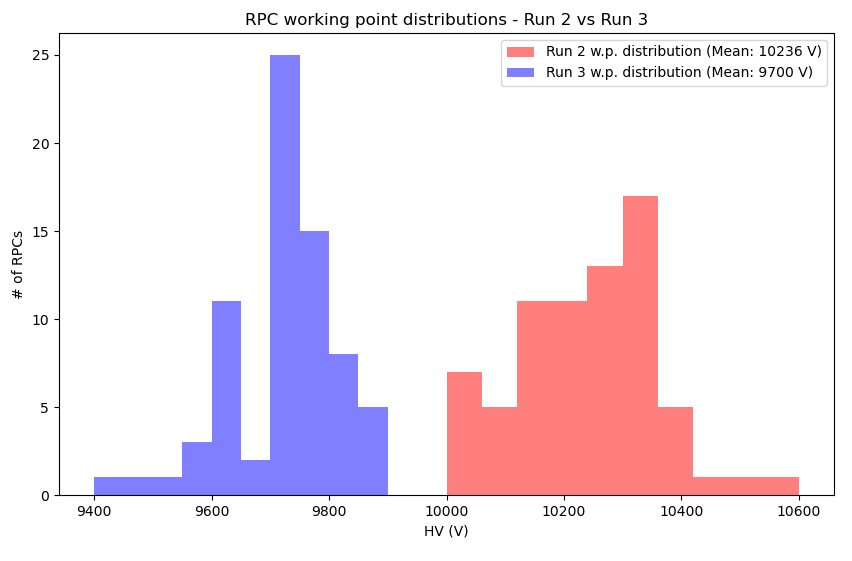}
    \caption{HV working point distribution in Run 2 and Run 3. The average MID HV in Run 3 is about 500 V lower with respect to Run 2.}
    \label{fig:HVdistr}
\end{figure}

The final HV working point was determined by selecting the HV value where the efficiency curve reaches the plateau, plus an additional 100 V. Following this fine-tuning process, the average MID HV in Run 3 is approximately 500 V lower compared to Run 2 (see Fig. \ref{fig:HVdistr}).

\begin{figure}[htb]
    \centering
    \includegraphics[width=10cm,height=5cm]{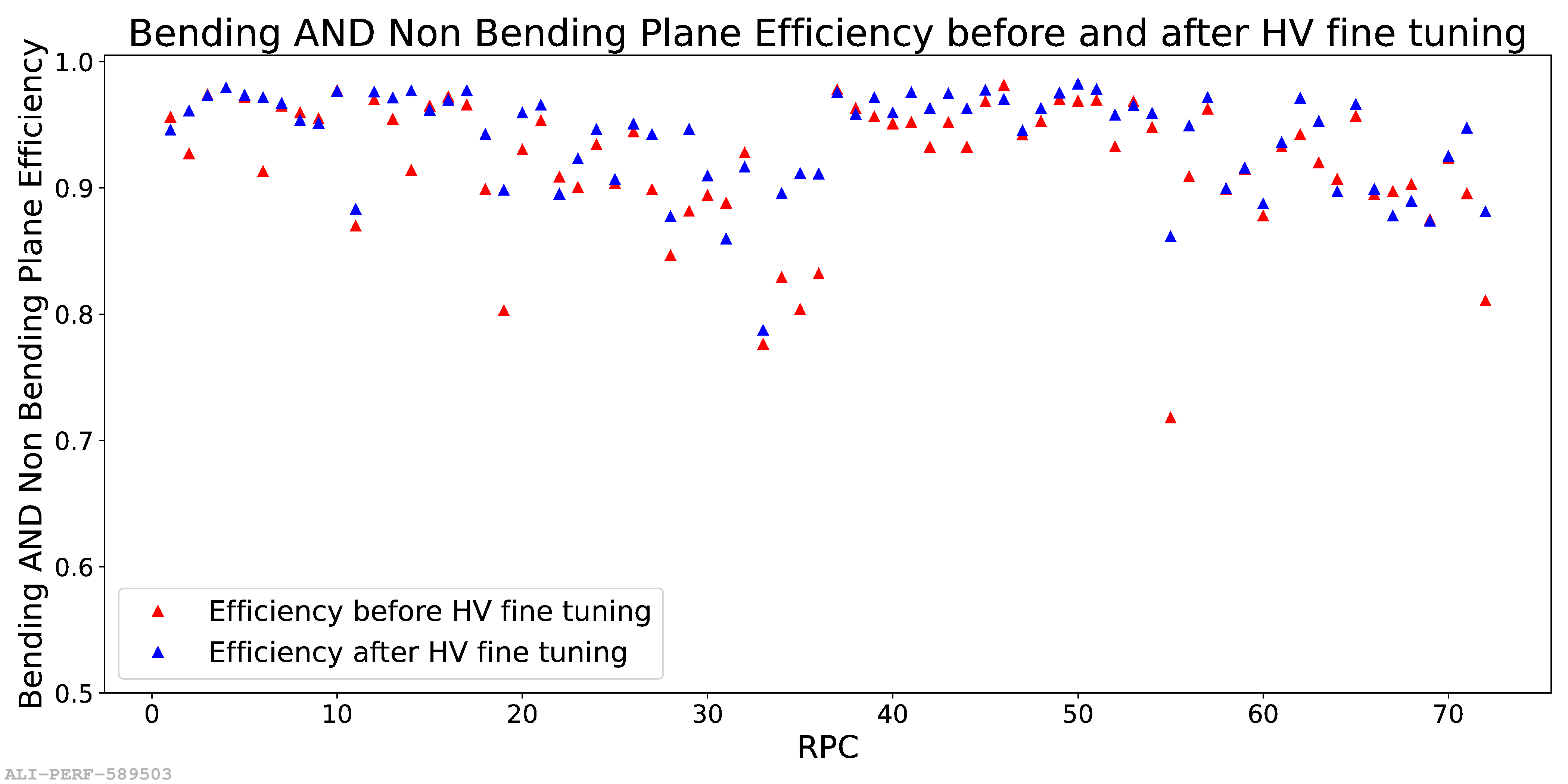}
    \caption{Efficiency at working point of the 72 RPCs before and after the HV w.p. fine tuning.}
    \label{fig:averagerpceff}
\end{figure}

The efficiency of the 72 RPCs at the working point before and after the HV fine-tuning is shown in Figure \ref{fig:averagerpceff}. As expected, a clear improvement in efficiency is observed for all RPCs.
 
\newpage

\subsection{Efficiency stability in Run 3}

\begin{figure}[htbp]
\centering
\includegraphics[width=.7\textwidth]{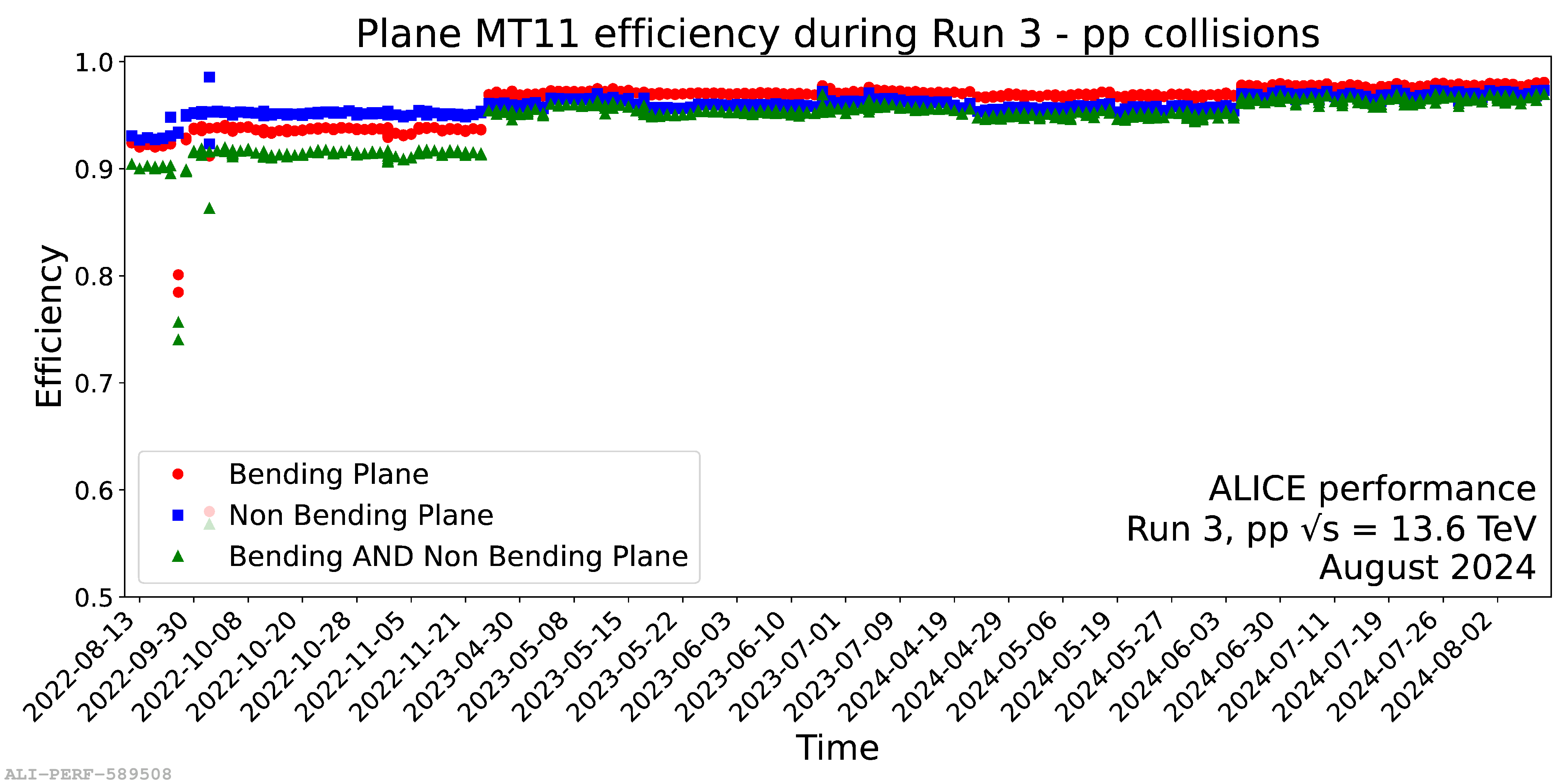}
\qquad
\includegraphics[width=.7\textwidth]{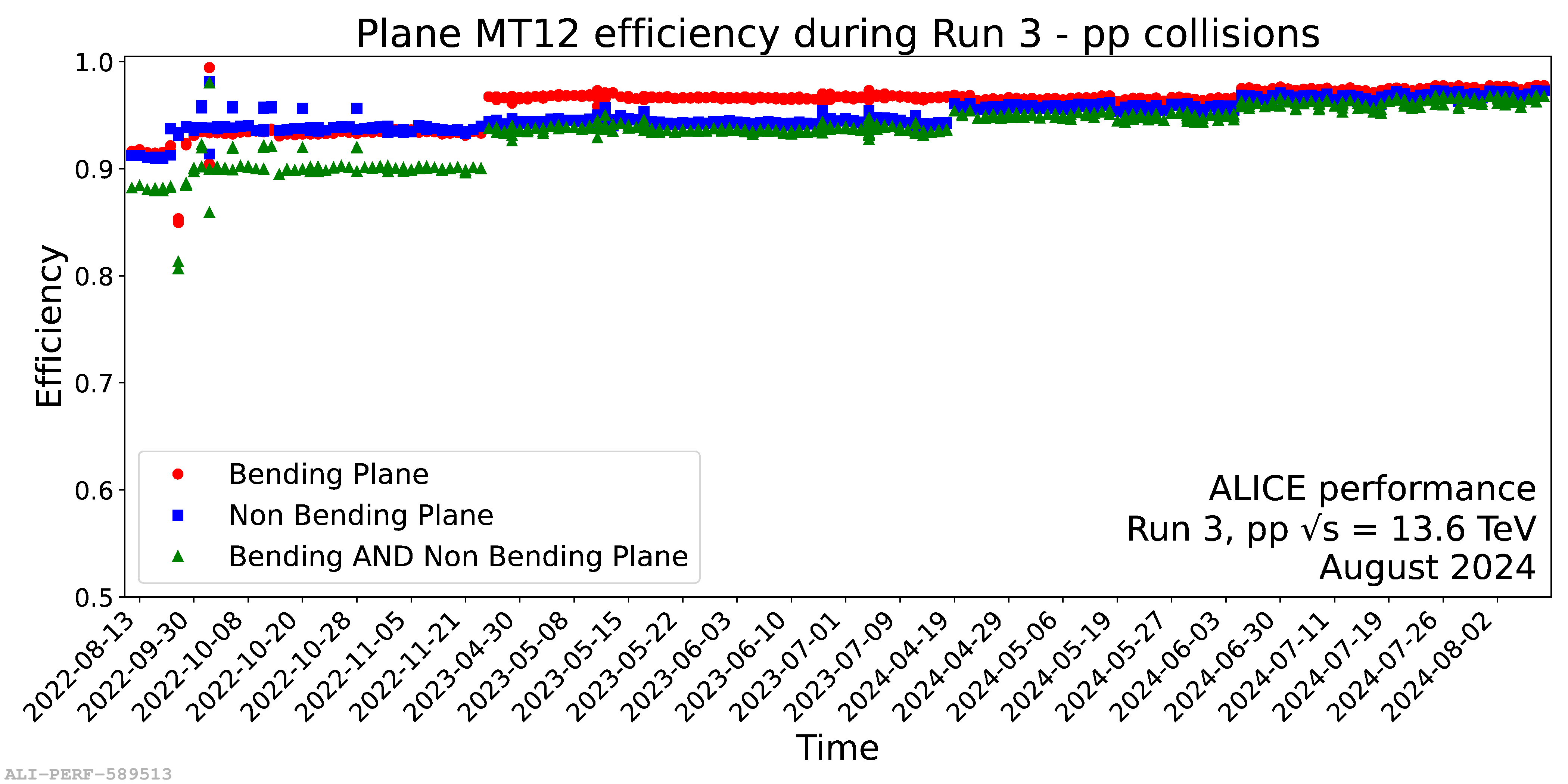}
\caption{Average efficiency as a function of time during Run 3 (p-p collisions only), of two of the four MID planes, i.e. MT11 and MT12. BP, NBP and the AND of the two are shown separately. \label{fig:run3eff}}
\end{figure}

The average efficiency of the MID has been continuously monitored throughout the Run 3 data-taking period. \\
As an example, the average efficiency over time for two of the four MID planes, specifically MT11 and MT12, during Run 3 is shown in Figure \ref{fig:run3eff}. Each plot contains three curves representing the Bending Plane, the Non-Bending Plane, and the AND of the two.
%In this case p-p collisions only are shown, since the fine tuning of the evaluation of efficiencies for Pb-Pb collisions is still in progress. 
As shown in the plots, the efficiency has improved each year from 2022 to 2024, thanks to various hardware (e.g., replacement of faulty RPCs and front-end electronics boards) and software (e.g., enhancements in the reconstruction algorithm) upgrades.  The efficiency has consistently remained above 95$\%$.

\subsection{Dark currents}

In addition to efficiency monitoring, the dark currents of the RPCs were also continuously observed during the data-taking period. The dark current is defined as the amount of current flowing through the RPC when it is switched on at nominal HV and fully operational in the absence of LHC beams (for instance, when ALICE is recording cosmic data). \\
It was observed that, compared to Run 2, the average MID dark current has decreased by a factor of 5, measuring 1 $\mu A$ in Run 3 compared to 5 $\mu A$ in Run 2, thanks to the new operational mode.

\subsection{First performance for quarkonium physics}

\begin{figure}[htb]
    \centering
    \includegraphics[width=8cm,height=6.5cm]{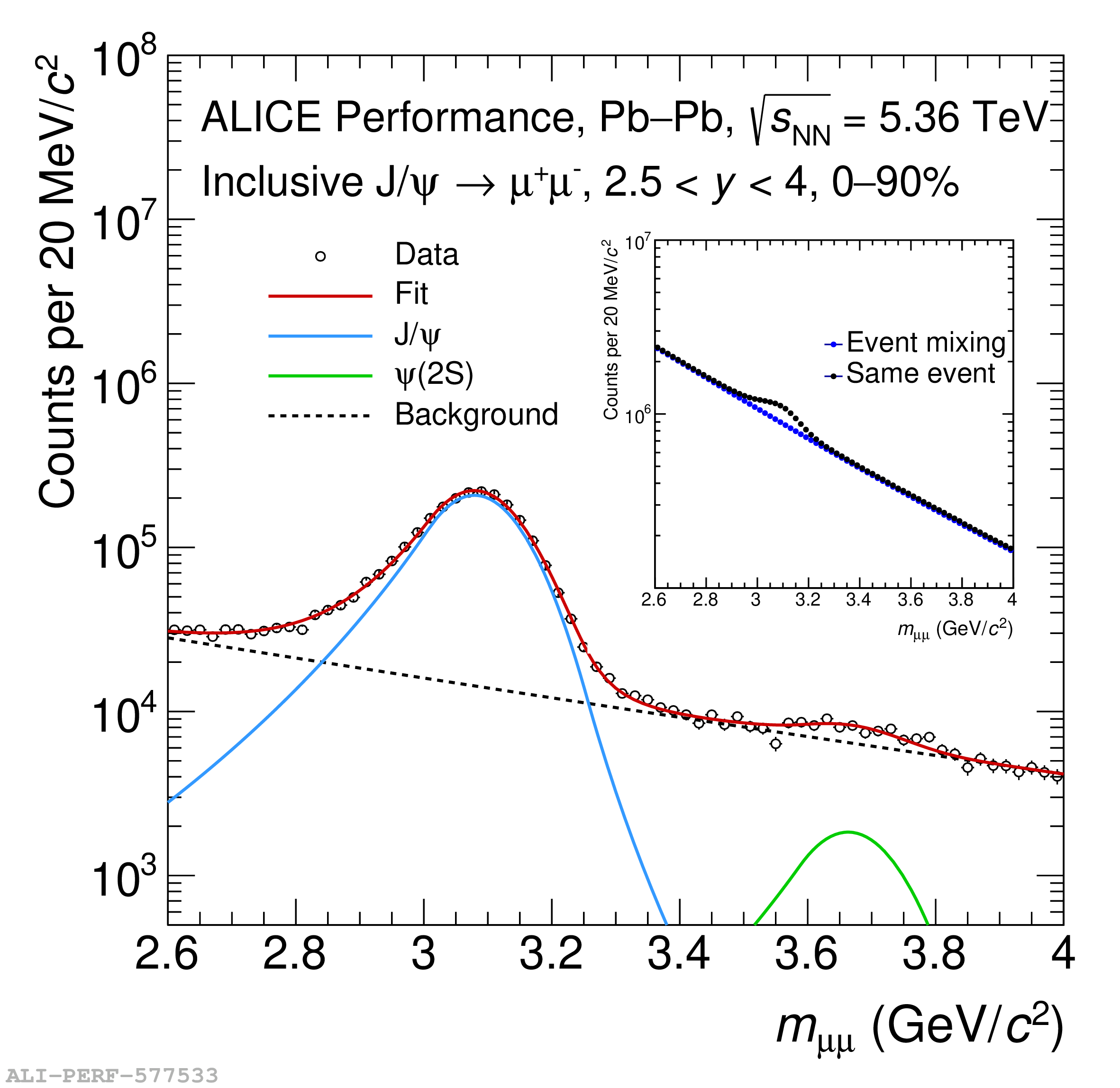}
    \caption{Invariant-mass distributions of quarkonia at forward rapidity in Pb-Pb collisions at $\sqrt{(s_{NN})}$ = 5.36 TeV. The insert plot presents the invariant-mass spectrum of \textit{J/$\psi$} along with the mixed-event background. The main plot displays the invariant-mass and corresponding fits for \textit{J/$\psi$} and \textit{$\psi$(2S)} based on the invariant-mass spectrum shown in the insert plot, after subtracting the mixed-event background.}
    \label{fig:jpsi}
\end{figure}

The first performance plot for quarkonium physics in Run 3 is shown in Fig. \ref{fig:jpsi}. The plot shows the invariant-mass distribution of opposite sign muon pairs around the \textit{J/$\psi$} and \textit{$\psi$(2S)} mass at forward rapidity in 0 - 90$\%$ central Pb-Pb collisions. Tracks reconstructed in the ALICE muon tracking chambers and matched with MID track segments were used, with a $p_T$ between 0 and 30 GeV/c, and the background is evaluated from mixed events. The insert plot presents the invariant-mass spectrum of \textit{J/$\psi$} along with the mixed-event background for Pb-Pb collisions at $\sqrt{(s_{NN})}$ = 5.36 TeV. The main plot displays the invariant-mass and corresponding fits for \textit{J/$\psi$} and \textit{$\psi$(2S)} based on the invariant-mass spectrum shown in the insert plot, after background subtraction.

 \section{Conclusions}
 \label{Conclu}
 
During LS2 the ALICE MID was successfully upgraded for Run 3 to cope with the higher interaction rates. Since the start of Run 3, the detector has been operating smoothly in both p-p and Pb-Pb collisions, with a low percentage of bad runs, primarily caused by readout failures. A run is classified as BAD for MID when a portion of the detector corresponding to 30 Local Boards (out of 234) fails to send data, resulting in compromised track reconstruction in the Muon Spectrometer.
In Run 3, the overall percentage of bad runs for p-p collisions is 7$\%$, with the majority occurring during the 2022 data-taking period, and showing a decreasing trend each year. For Pb-Pb collisions, the percentage of bad runs is significantly lower at only 1.4$\%$. The new FEE enables operation at a lower HV, reduced by 600 V compared to Run 2, and the average dark current is lower by about a factor 5. The HV scan was successfully performed to allow the fine tuning of the new HV working points. Continuous improvements in both hardware and software have resulted in a steady increase in the average efficiency each year, with the majority of RPCs achieving an efficiency greater than 95$\%$. Further studies are ongoing at the Local Board efficiency level to enhance overall performance. In conclusion, the performance for \textit{J/$\psi$} and \textit{$\psi$(2S)} measurements has demonstrated promising results, further confirming the effectiveness of the MID upgrades.



\end{document}